\documentclass[fleqn,usenatbib]{mnras}
\usepackage{newtxtext,newtxmath}
\usepackage[T1]{fontenc}
\DeclareRobustCommand{\VAN}[3]{#2}
\let\VANthebibliography\thebibliography
\def\thebibliography{\DeclareRobustCommand{\VAN}[3]{##3}\VANthebibliography}
\usepackage{graphicx}	% Including figure files
\usepackage{amsmath}	% Advanced maths commands
\usepackage{float}
\usepackage{tikz}
\usepackage{booktabs}
\usetikzlibrary{shapes.geometric, arrows}
\tikzstyle{startstop} = [rectangle, rounded corners, text centered, draw=black, fill=red!30]
\tikzstyle{database} = [cylinder, cylinder uses custom fill, cylinder body fill=yellow!30, cylinder end fill=yellow!30, shape border rotate=90, aspect=0.25, draw=black, text centered]
\tikzstyle{input} = [trapezium, trapezium left angle=70, trapezium right angle=110, text centered, draw=black, fill=blue!30]
\tikzstyle{decision} = [diamond, text centered, draw=black, fill=green!30]
\tikzstyle{process} = [rectangle, text centered, draw=black, fill=orange!30, minimum width=3cm, minimum height=1cm]
\tikzstyle{arrow} = [thick, ->, >=stealth]

\def\logg{$\log~g$}
\def\teff{T$_{\rm{eff}}$}
\def\vsini{$v \sin i$}
\def\vsinitheo{$\langle v \sin i \rangle_{\rm sim}$}
\def\vsiniFT{$\langle v \sin i \rangle_{\rm FT}$}
\def\vsinifinal{$\langle v \sin i \rangle_{\rm final}$}
\def\vsiniline{$\langle v \sin i \rangle_{\rm line}$}
\def\sigmafit{$\sigma_{\rm fit}$}
\def\sigmanoise{$\sigma_{\rm noise}$}
\def\sigmac{$\sigma_{\rm c}$}
\def\AxP{\AA/pix}
\def\kps{~km~s$^{-1}$}
\def\HeI{He {\sc{i}}}
\def\LHeIOne{$\lambda$4471}
\def\lamtheo{$\lambda_{\rm theo}$}
\def\AllLamdas{$\lambda \lambda$4387, 4471, 4713, 4921, 5015, 5047, 5875 and 6678}

\title[\vsini~via Fourier Transform in BeSOS]{Automatic algorithm to obtain \vsini~values via Fourier Transform in BeSOS database}
\author[Solar et al.]{
Martín Solar,$^{1,2}$\thanks{E-mail: martin.solar@postgrado.uv.cl}
Catalina Arcos,$^{1}$
Michel Curé,$^{1}$
Ronaldo S. Levenhagen$^{3}$
and Ignacio Araya$^{4}$ \\
$^{1}$Instituto de Física y Astronomía, Facultad de Ciencias, Universidad de Valparaíso, Gran Bretaña 1111, Valparaíso, Chile\\
$^{2}$Astronomical Observatory Institute, Faculty of Physics, Adam Mickiewicz University, ul. Słoneczna 36, 60-286 Poznań, Poland\\
$^{3}$Departamento de Física, Universidade Federal de Sao Paulo, Rua Prof. Artur Riedel, 275, 09972-270, Diadema, SP, Brazil\\
$^{4}$Centro de Investigación DAiTA Lab, Facultad de Estudios Interdisciplinarios, Universidad Mayor, Alonso de Córdova 5495, Santiago, Chile}

\date{Accepted XXX. Received YYY; in original form ZZZ}
\pubyear{2021}

\begin{document}
\label{firstpage}
\pagerange{\pageref{firstpage}--\pageref{lastpage}}
\maketitle

\begin{abstract}
Be stars are found to rotate close to their critical rotation and therefore they are considered as an important laboratory of study for stellar rotation. In this context, we obtain the projected rotational velocity of a sample of classical Be Southern stars in the BeSOS database via Fourier Transform in an automated way for several absorption lines at different epochs. A Gaussian profile is fitted to eight observed photospheric \HeI~lines in order to select automatically the spectral signal given by areas under the curve of 95.45~\%, 98.75~\% and 99.83~\% from the profile to obtain \vsini~via Fourier Transform technique. The values obtained are in global agreement with the literature. Analysing only one line is not enough to set the \vsini~value, depending on the line the value in most cases are underestimated with respect to \LHeIOne. When gravity darkening effects are including, apparent values increases by $\sim10$~\%. The resolution of the instrument PUCHEROS used for BeSOS spectra ($R \sim 17\,000$) constrain a theoretical lower bound possible at \vsini~$\sim 100$~\kps. The procedure has limitations using a linear limb-darkening function with $\varepsilon = 0.6$ for classical Be stars rotating close to the break-up velocity without gravity-darkening corrections, which can't be negligible. Previous works measure \vsini~values using just one spectral line and here we demonstrate that with more lines the results can varies. This could be due to the photospheric distribution of atomic transitions on classical Be stars.
\end{abstract}


\section{Introduction}

The estimation of stellar rotation velocities from spectral line widths dates 
back to the early work by \cite{Abney1877}. Today, there are multiple methods at 
our disposal to measure the projected rotation velocity (\vsini), but systematic 
differences among them are still challenging to deal with. \cite{Shajn1929} suggested a simplified 
graphical model, without limb-darkening, which turned out to be a standard reference 
through decades \citep{Collins-Truax1995ApJ...439..860C}. 
Later, Slettebak published several works \citep{Slettebak-Howard1955ApJ...121..102S,Slettebak1954ApJ...119..146S,Slettebak1955ApJ...121..653S,Slettebak1956ApJ...124..173S} aiming to measure \vsini~using this feature. These works were further revised 
and a new standard system of rotating velocity stars across the HR diagram was elaborated 
\citep{Slettebak1975ApJS...29..137S}. These works gave rise to systems of standard stars of rotation 
velocity, subsequently adopted in calibrations of the full width at half-maximum (FWHM), or 
even the equivalent width, against \vsini. 

Another class of methods, relying on the Fourier transform (FT) of line profiles, to derive 
\vsini~ from the first measured roots, were first proposed by \cite{Carroll1928,Carroll1933_a,Carroll1933_b} 
and are extensively used in the literature \citep{Ilin1979,Jankov1995,Reiners2002,Reiners2003,Royer2005,SimonDiaz2006,SimonDiaz2007,Diaz2011,RamirezAgudelo2013}. 
When compared to Shajn's simple graphical model, these methods have the advantage of being free of external calibrations and allowing the identification of extra broadening 
mechanisms and even differential rotation provided the line profiles have a high signal-to-noise 
(S/N) relation \citep{Gray1977ApJ...211..198G,Bruning1981ApJ...248..274B,Reiners2002}. 
More sophisticated methods of this kind, like that proposed by \cite{Diaz2011}, employ the 
cross-correlation function (CCF) of the profile against a zero-rotation template and apply 
the FT of the CCF maximum to measure \vsini.

In the present work we intend to apply the classical FT method, specifically to the study of Be stars.
Classical Be stars (CBes) are B-type main sequence (MS) stars holding, at least once in their 
lives, a geometrically thin equatorial disc rotating in a Keplerian way \citep{Rivinius2013}. 
The nature of the disc formation process is not completely understood, but stellar rotation 
plus additional mechanisms, are proposed to play an important role \citep{Baade2016}. As CBes 
rotate close to their critical velocity limit \citep{Porter2003}, the rotation kernel dominates over
other broadening mechanisms. Additionally, these stars are generally 
deformed by rotation, leading to an oblate spheroidal shape. As a result, the temperature and
gravity in the surface has a strong dependency with the stellar latitude. Regions closer to the
equator have the greater velocities, but are at the same time the coldest ones, so the flux
contribution from these regions to the spectrum are lesser than for higher latitudes. This effect is
known as gravity-darkening (GD) and was first reported by \cite{vonZeipel1924_a,vonZeipel1924_b}.

Until today, there is not a variant of the FT method capable of estimating \vsini~ and deal with GD
at the same time. In a recent work, \cite{Fremat2005} performs a sophisticated analysis among 
rotating models affected by both gravity-darkening (GD) and rotational distortion, and parent non-rotating 
counterpart model spectra, to yield ``true'' \vsini~values, but this analysis does not involve FT's.

High stellar rotation impacts directly the evolution of stars. As the star rotates near the break-up 
velocity, its shape becomes oblate and the gas pressure goes down at the equatorial regions. This 
leads to a decrease in luminosity and a potential increase in its life along the main sequence.
Also, the meridional circulation induced by rotation, known as Eddington-Sweet currents, brings the
elements formed deep inside the star to its surface, and changing the chemical content of its 
photosphere \citep{Maeder2010}.

Concerning CBes, helium lines are more dominated by the rotational kernel than other broadening mechanisms such as intrinsic, thermal, turbulent, or Stark, supporting their use. Also, there are essential benefits to using FTs instead of directly comparing an observed line profile with stellar atmosphere models because the line shape is more prominent in the Fourier domain than in the wavelength domain for the reason that includes the whole feature of the spectral signal. Additionally, Gaussian noise is more easily distinguished from the signal \citep{Smith1976}. Usually, only the \HeI~\LHeIOne~is used to measure \vsini~in CBes because the depth formation is higher than Balmer lines and in this work we use seven more \HeI~photospheric lines for comparison purposes \citep{Mihalas1964, Omara1972, Neiner2002, Vinicius2006}. In the case of rapid rotators (100–400 km s$^{-1}$) GD effects given by temperature inhomogeneity on surfaces cause considerable variations in the strength and profile shape of spectral lines on the stellar disc \citep{Preston2019}. In this
work, we use a code by \cite{Levenhagen2014} to calculate line profiles assuming GD, and we 
estimate the ``true'' \vsini\, values from the interpolation of corrections in temperature, gravity and \vsini\, by \cite{Fremat2005}.

Multiple works have been done in obtaining the projected rotational velocity in stars, however, in most of them there are no mathematical model that quantify an automatic algorithm to select the signal of an absorption line broadened by Doppler effect due to rotation, and for instance, \cite{SimonDiaz2014} select manually the midwave and continuum of the absorption line to compute the FT. New telescopes and surveys are challenging the form to handle astronomical data \citep{York2000,Skrutskie2006,Gaia2021} and in this work we introduce a procedure using Gaussian fits for the automatic determination of \vsini~based on the FT method since observations growing constantly and it is essential to develop a mechanism capable to deal with this. We performed tests of this algorithm using Monte Carlo (MC) simulations of synthetic spectra affected by different noise levels. The algorithm was applied to ``Be Star Observation Survey'' (BeSOS) spectra and compared with results from other works. 

The outline of the paper is as follows. In Section 2, we present the BeSOS database used in this work. In Section 3, the methodology describes how we propose to obtain \vsini~values using the FT technique. In Section 4, we apply the MC simulations to evaluate the range of applicability of the method. In Section 5, we show the results obtained for BeSOS spectra. Finally, in Sections 6 and 7, the discussion and conclusions are presented, respectively.

\section{BeSOS database}
In order to test the developed algorithm, we made use of BeSOS spectra \citep{Arcos2018}, a database which is dedicated to analyse and study the variability of Southern Be stars through optical spectroscopic follow-up. BeSOS contains more than 60 targets (with a limiting visual magnitude of V=9) observed between 2012 and 2015, and more than 300 spectra in total with signal-to-noise ratio (SNR) $\sim$ 100. All spectra are observed with the same instrument, a mid-Resolution ($R\sim17\,000$) Echelle spectrograph, PUCHEROS \citep{Vanzi2012} covering the visible spectral range ($4260-7300 \ \rm \mathring{A}$) and mounted at the 50cm ESO Telescope in the ObsUC located in Santiago, Chile. The data were processed using the CERES pipeline \citep{Brahm2017} and the continuum normalisation was made using IRAF packages \citep{Tody1993}. For more details, we refer to the main BeSOS paper \cite{Arcos2018} and references therein. The spectra are available to download in the BeSOS website\footnote{\url{besos.ifa.uv.cl/}}. Also, the website provides information about the projected rotational velocity values obtained by fitting rotational convoluted stellar atmosphere models \citep{Kurucz1979, Hubeny1995} to the observed line profile \HeI~\LHeIOne.

\section{Methodology}

A spectral line profile can, in principle, be regarded as the convolution between an intrinsic intensity profile and a rotation kernel which is a function of limb-darkening (LD) coefficients and wavelength shifts $\rm \delta = \alpha / \alpha_L$ \citep{Gray2005}, where we denote for simplicity $\rm \alpha = \Delta \lambda$ to be the wavelength displacement coordinate and $\rm \alpha_L = \Delta \lambda_L$ to be the limit wavelength range of the line profile. Depending on the polynomial fitting that we choose to reproduce the intensities of a spectrum, we will obtain different rotation profiles \citep{Kopal1950}. In order to apply this technique to the observed line profiles, the broadening mechanism should be  
dominated by rotation.

Departing from the most simple polynomial line intensities emerging from all stellar directions ($I$), i.e. a linear LD law \citep{Carroll1933_a,Carroll1933_b}, and assuming spherical stars and rigid rotation we can write the intensity as follows:

\begin{equation}\label{eq1}
I(\cos{\theta}) = I(1)\,[1-\varepsilon\,(1-\cos{\theta})],
\end{equation}

\noindent where the disc's centre is at $\cos \theta = 1$ and the limb given by $\cos \theta = 0$, and $\varepsilon \in [0,1]$ is the LD coefficient. By this way the rotation profile can be written as \citep{Gray2005}:

\begin{equation}\label{eq2}
G(\delta) = c_1\,(1 - \delta^2)^{1/2} + c_2\,(1 - \delta^2),
\end{equation}

\noindent where the linear coefficients are given by:

\begin{equation}\label{eq3}
c_1 = \frac{2}{\pi}\,\left( \frac{1-\varepsilon}{1-\varepsilon/3} \right) \,\,\, \text{and} \,\,\,   
c_2 = \frac{1}{2}\,\left( \frac{\varepsilon}{1-\varepsilon/3} \right).
\end{equation}

The FT of the rotational profile is treated in deep by \cite{Levenhagen2014} for an analogous problem involving different expressions. A linear expression remains, namely:

\begin{equation}\nonumber
\mathscr{F}(G(\alpha)) = \int_{-1}^{1} G(\alpha)\,\cos{(2 \pi \sigma \alpha)}\,d\alpha
\end{equation}

\begin{equation}\label{eq4}
\,\,\,\,\,\,\,\,\,\,\,\,\,\,\,\,\,\,\,\,\,\,\,\,\,\,= c_1 \,\frac{J_1(2 \pi \sigma)}{2 \sigma} + c_2 \,\frac{1}{\pi^2 \sigma^2}\,\left( \frac{\sin{2 \pi \sigma}}{2 \pi \sigma} - \cos{2 \pi \sigma} \right),
\end{equation}

\noindent where $\sigma$ is the Fourier coordinate (in units of cycle/\AA) associated to the wavelength $\alpha$ coordinate. Assuming $\varepsilon=0.6$, the first root of the FT occurs at $\alpha_{\rm L} \sigma_{0} = 0.66$. Dividing this root by the first root of the FT of the observed profile will result in the $\rm \alpha_{\rm L}$ estimate, which is proportional to \vsini. It is commonly assumed a linear LD coefficient of $\varepsilon=0.6$ in spectral synthesis routines and stellar rotational calculations and it is worth emphasising the strong dependence of $\varepsilon$ with temperature, gravity, metallicity, and wavelength \citep{Wade1985}.

\begin{figure*}
\resizebox{17.5cm}{!}{\includegraphics{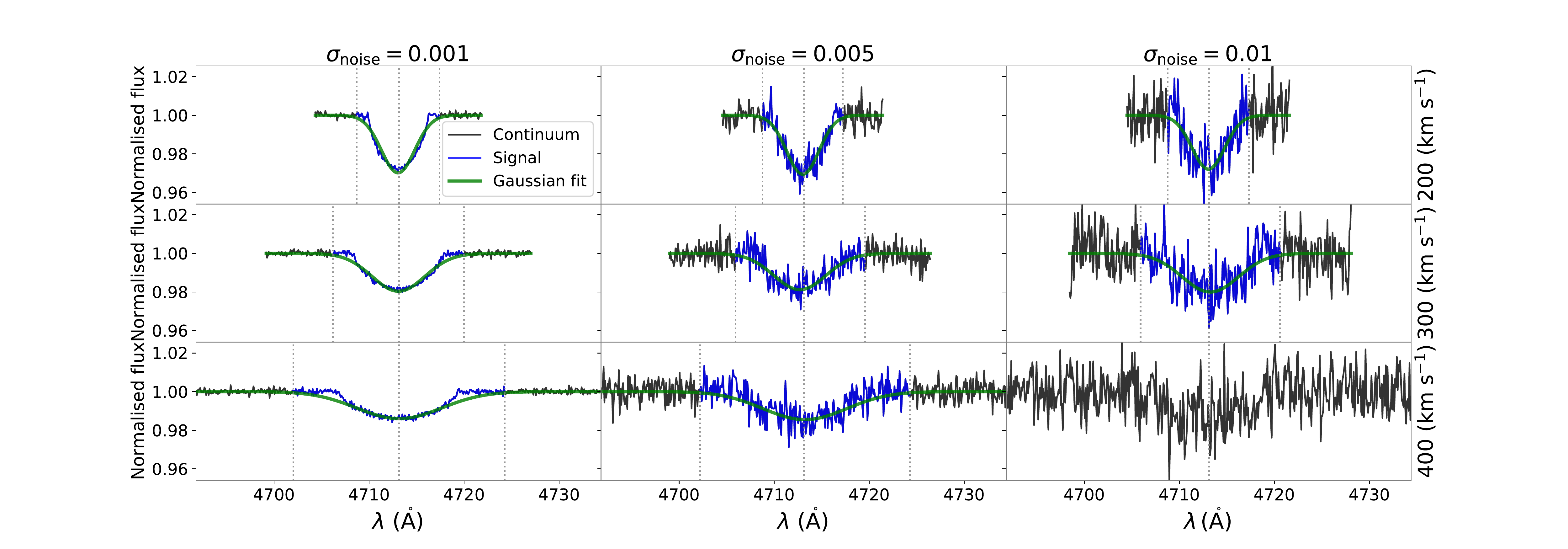}}
    \caption{Grid of simulated line profiles with noise and rotational kernels for 2.5~\sigmafit. The normalised flux is as a function of the wavelength. The solid green curve represents the best Gaussian fit, the solid blue lines are the selected spectral signal window and the solid black lines represent the continuum. These three sections are indicated by the vertical dotted grey lines. Left panels: \sigmanoise~$=$~0.001. Middle panels: \sigmanoise~$=$~0.005. Right panels: \sigmanoise~$=$~0.01. Top panels: \vsinitheo~$=$~200~\kps. Middle panels: \vsinitheo~$=$~300~\kps. Bottom panels: \vsinitheo~$=$~400~\kps.}
    \label{fig2.1}
\end{figure*}

We employ the expression found by \cite{Dravins1990} for the first zero of eq.~\ref{eq4} to derive \vsini~using our automated algorithm. It is a power series of the LD coefficient $\varepsilon$ with precision of a unit on the third decimal place:

\begin{equation}\label{eq5}
\alpha_{\rm L} \, \sigma_{0}=0.610 + 0.612\,\varepsilon + 0.027\,\varepsilon^{2} + 0.012\,\varepsilon^{3} + 0.004\,\varepsilon^{4}\, ,
\end{equation}

\noindent where $\alpha_{\rm L} = \Delta \lambda_{\rm L}=(\lambda_{\rm theo}/c)\,$ \vsini~represents the maximum broadening that takes place at the stellar limbs (\lamtheo~is the central wavelength of a \HeI~spectral line) and $\sigma_{0}$ is the first zero of the FT frequency domain. \vsini~values obtained by this method are highly dependent of how noisy and how larger is the selected range (spectral signal window) of the absorption line, in which in order to study this behaviour, we simulate an \HeI~line $\lambda_{\rm theo} = $~4713.14~\AA~considering three different noise levels (from a pseudo-random Normal distribution), \sigmanoise~$=$~0.001, 0.005 and 0.01 (dimensionless), and three different theoretical projected rotational velocity values, \vsinitheo~$=$~200, 300 and 400\kps~(see Fig.~\ref{fig2.1}).

For the selection of the spectral signal window, a Gaussian profile is fitted to the absorption line starting with initial parameters of an amplitude of 0.025 and a standard deviation computed in a range of $\pm$~5~\AA~with respect to \lamtheo. The best parameters are found using a standard non-linear least squares calculation, and hence, the optimal curve is computed. In this case, the information inside $\pm k \, \sigma _{\rm fit}$ (dimensionless) window with respect to \lamtheo~is selected, being \sigmafit~the standard deviation of the optimal Gaussian fit. In this work we use $k =$~2.0, 2.5 and 3.0. The data outside this window is not considered and is defined as continuum. This procedure computes \vsini~selecting automatically the spectral window region to be used in the FT method in any absorption line (\HeI~for our work). Nevertheless, we still need to quantify the obtained \vsini~value as a function of the quality of the observed absorption line, i.e. noisy spectra can yield significant errors, regardless of the parameters to be obtained. For this reason, in this work we propose five criteria described in Table~\ref{tab:criteria} (The sixth criterion is explained in detail in section \ref{sect4}).

\begin{table}
    \begin{tabular}{    p{3.6cm}  p{4.1cm} }
        \toprule
\textbf{Criteria}  & \textbf{Description}   \\\midrule
\textbf{1st criterion:} A Gaussian profile is successfully fitted given the initial parameters (see text for initial parameters values). 
& If the observed line profile is undetectable with respect to the continuum, then the numerical fit it is going to fail at the moment to find the optimal curve. \\\hline

\textbf{2nd criterion:} An absorption line exists when the fitted Gaussian profile has a peak less than 1 (simulated normalised flux).
& An atomic transition must be in absorption to trace the atmospheric surface, and sometimes, this line can be contaminated with information from the disk.  \\\hline

\textbf{3rd criterion:} The centre of the Gaussian fit is not shifted more than $k$~\sigmafit~with respect to $\lambda_{\rm theo}$.
& In some cases there is no absorption line in $\lambda_{\rm theo}$ and numerically another transition line close to it is found, as is in the case of \HeI~\LHeIOne~and MgII $\lambda$4481. Also, considering the radial velocity displacement from $\lambda_{\rm theo}$, the quantity $k$\sigmafit~leaves a considerable range of error to find the absorption line.\\\hline

\textbf{4th criterion:} The standard deviation of the continuum \sigmac~(dimensionless) is lesser than the amplitude of the Gaussian fit.
& The goal of this criterion is differentiate if the continuum is too noisy with respect to the line profile. \\\hline

\textbf{5th criterion:} The dispersion of the Gaussian fit is higher than 5 times the step of the wavelength $\lambda$.
& To quantify the quality of the signal, there should be at least 5 bins in one \sigmafit~from the Gaussian fit. This tell us that the line has significantly information inside the Gaussian profile. \\\hline

\textbf{6th criterion:} The standard deviation of the continuum in observations is lesser than the obtained from MC simulations. 
& Given 10~000 MC simulations for a grid of simulated \sigmanoise~and \vsinitheo, if the \sigmac~in observations is lesser than median of the simulated ones for a specific velocity, then the absorption line is considered to measure \vsini. \\
        \bottomrule
    \end{tabular}
    \caption{Criteria proposed to check if an observed absorption line is considered as a suitable signal.}
    \label{tab:criteria}
\end{table}

If the line satisfies these five criteria, a \vsini~value is computed by the FT method, otherwise, the data is not utilised. It is expected that as noise increases, the projected rotational velocity obtained should be less reliable and to obtain a correct velocity. The criteria try to generalise a well automatic quantification for any absorption line and it  depends from the instrument used. In this work, the construction of simulated line profiles are given by as the same spectra resolution of Pontificia Universidad Catolica High Echelle Resolution Optical Spectrograph (PUCHEROS) and the Fiber-fed Extended Range Optical Spectrograph \citep[FEROS;][]{Kaufer1999} with wavelength steps of 0.089~\AxP~and 0.03~\AxP, respectively.

\begin{figure*}
\resizebox{17.5cm}{!}{\includegraphics{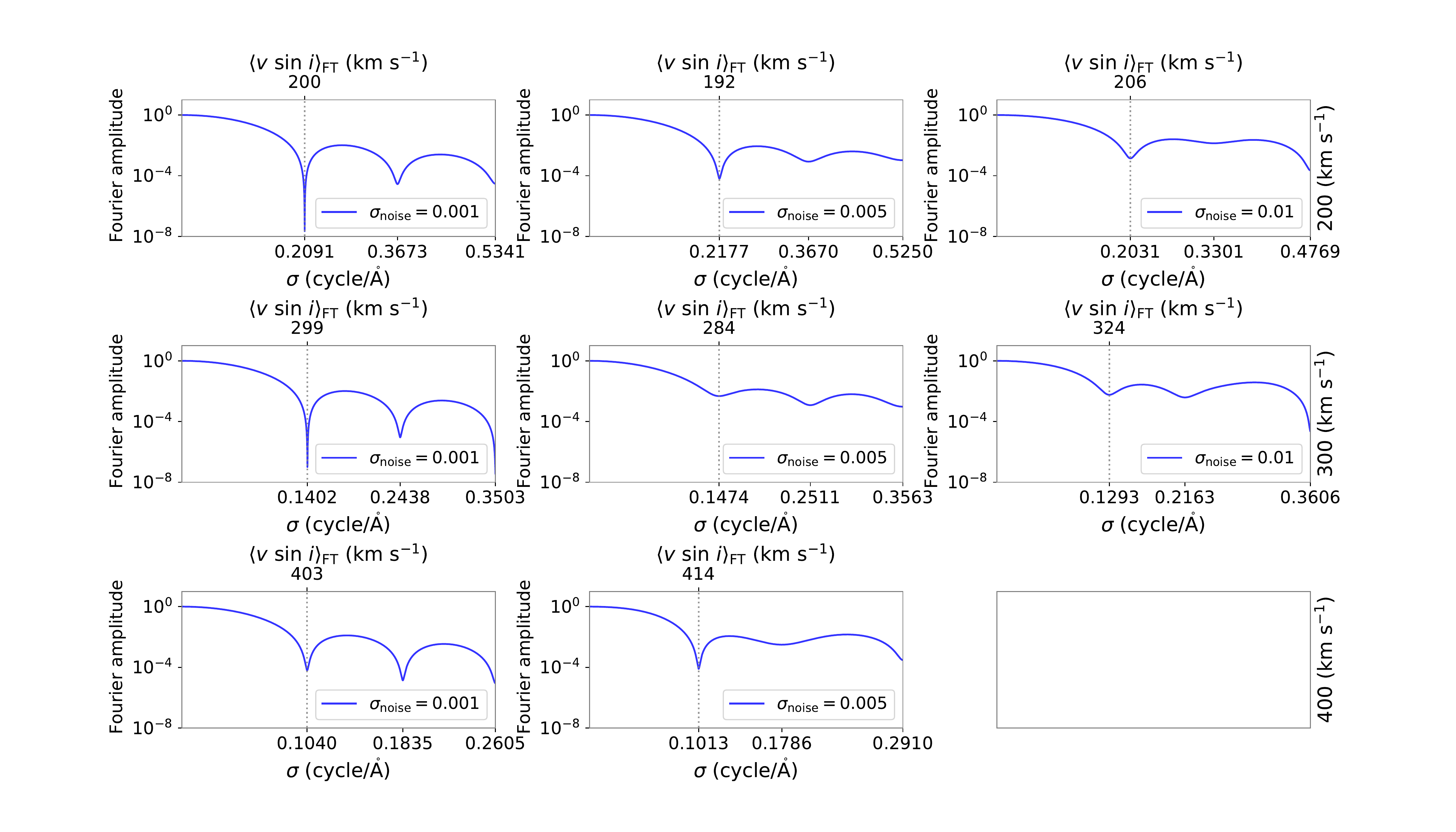}}
    \caption{FTs for spectral signal windows of $\pm 2.5\sigma_{\rm fit}$ for Fig.~\ref{fig2.1} represented by a solid blue curve. The subplots show the Fourier amplitude as a function of the Fourier domain $\sigma$ and their respective conversion of the first zero to the obtained velocity \vsinitheo, which is denoted by the vertical dotted grey lines. Left panels: $\sigma_{\rm noise}=0.001$. Middle panels: $\sigma_{\rm noise}=0.005$. Right panels: $\sigma_{\rm noise}=0.01$. Top panels: \vsinitheo$= 200$~\kps. Middle panels: \vsinitheo$= 300$~\kps. Bottom panels: \vsinitheo$= 400$~\kps. For the case of $\sigma_{\rm noise}=0.01$ and \vsinitheo$= 400$~\kps~(right bottom panel), FT is not computed because the line does not satisfy the fifth criterion.}
    \label{fig2.2}
\end{figure*}

For the simulated lines from Fig.~\ref{fig2.1}, the FT was applied to every spectral signal windows and velocities in which the results given by the first zero, represents the rotational speed \vsiniFT~obtained, showed in Fig.~\ref{fig2.2}. In this figure, it can be noted for all \vsinitheo, the velocities returned are not equal to the simulated one, specially as \sigmanoise~increase. The combination of high projected rotational velocity and high noise level ($\sigma_{\rm noise}=0.01$ and \vsinitheo~$=$~400~\kps) gives as result a non-fulfilment of one or more of the required criteria and no FT is computed (line dropped). Additionally to this, for \vsiniFT~there is no clear pattern for the resulted velocities so far and the numerical procedure described in this section has been defined for a single line as an example and results for multiple velocities and noise levels are described in next section using MC simulations.

\section{Monte Carlo Simulations}\label{sect4}

Simulated line profiles rotationally convoluted at different noise levels are generated to obtain \vsini~via FT transform with the aim to evaluate the reliability of the procedure. As a first example we assume that the intrinsic line profile of a star can be model by a Gaussian profile with parameters of dispersion $0.1$ and amplitude $0.5$, this profile is convoluted with a grid of rotational kernels from 100 to 600 in steps of 10\kps. In observations it is usual to obtain systematic variations due to instrumental errors and physical effects, among others. To add random noise to the simulated line profile (intrinsic + rotation) a grid of $\sigma_{\rm noise}$ is included from $0.001$ to $0.1$ in steps of $0.001$. 

\begin{figure*}
\resizebox{17.5cm}{!}{\includegraphics{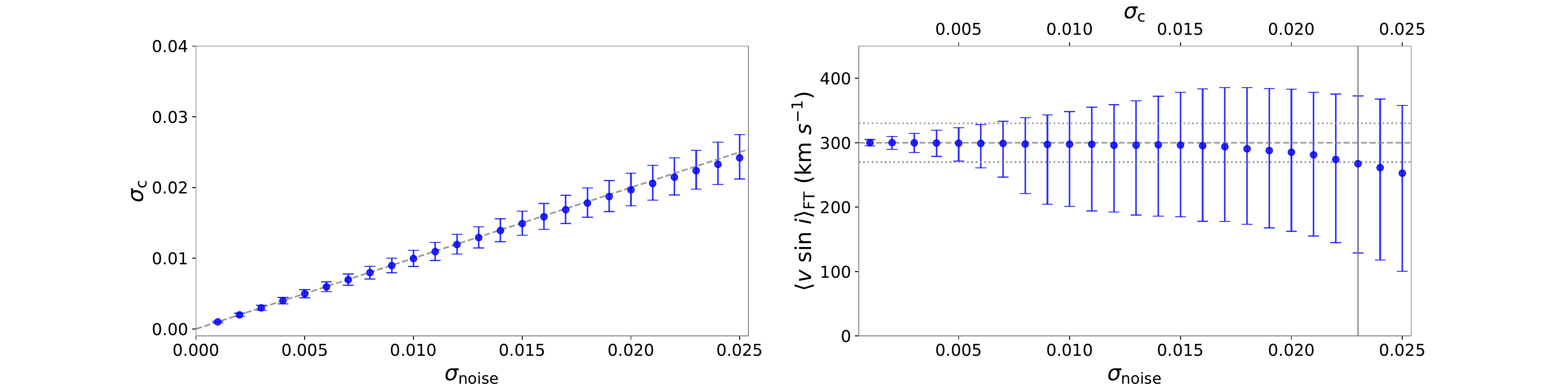}}
\resizebox{17.5cm}{!}{\includegraphics{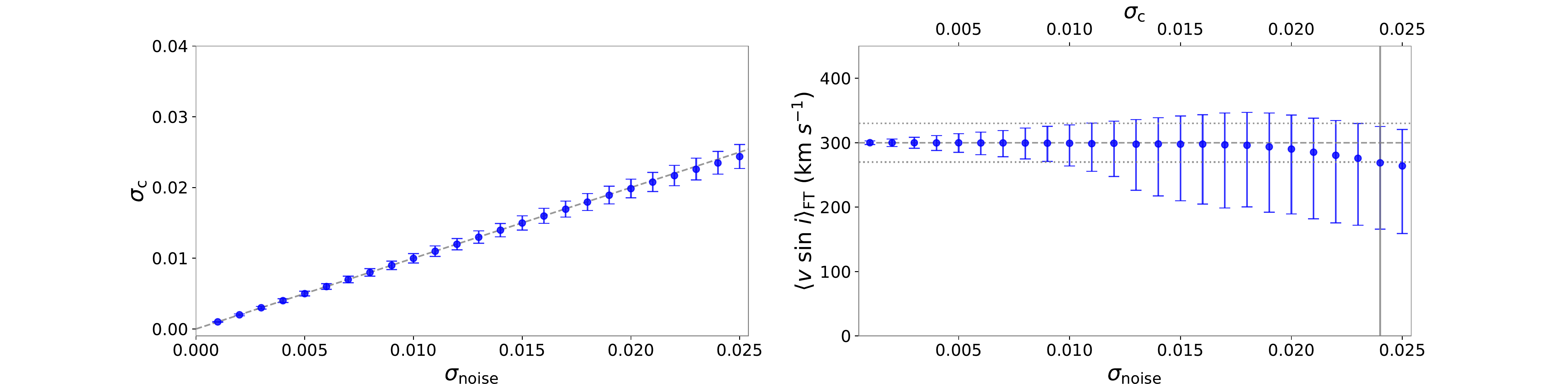}}
    \caption{Relation between critical $\sigma_{\rm noise}$ and $\sigma_{\rm c}$ for $2.5\sigma_{\rm fit}$, for 10~000 MC simulations, and with confidence intervals of $95\%$ (error bars from $2.5\%$ to $97.5\%$). Left panel: Standard deviation of the continuum as a function of the simulated noise added (blue dots). The 1:1 relation is represented by a dotted grey line. Right panel: Median of \vsiniFT~as a function of $\sigma_{\rm noise}=\sigma_{\rm c}$ (blue dots). The horizontal dashed grey line represents \vsinitheo$= 300 \ \rm km \ s^{-1}$ and the dotted grey dotted lines are $300 \pm 30 \ \rm km \ s^{-1}$. The vertical dotted grey line indicate $\sigma_{\rm c, crit}$. Top panels are simulations performed using PUCHEROS resolution and bottom panels for FEROS resolution.}
    \label{fig3.1}
\end{figure*}

In Fig.~\ref{fig3.1}, for 10~000 MC simulations and a window of 2.5~\sigmafit, it is shown the relation between $\sigma_{\rm c}$ (see Table \ref{tab:criteria}), fourth criterion and \vsiniFT~as functions of $\sigma_{\rm noise}$ for PUCHEROS and FEROS resolutions. The relation 1:1 in left panels evidence that $\sigma_{\rm c}$ trace totally $\sigma_{\rm noise}$, that is, assuming an observed line profile with Normal noise in the spectrum therefore is possible to obtain the dispersion parameter for the star computing the standard deviation of the normalised flux. The velocities estimated given by simulated line profiles of \vsinitheo$= 300 \ \rm km \ s^{-1}$ for the whole grid of $\sigma_{\rm noise}$ are presented in right panels. It can be noted that the \vsini~values decrease from $\sigma_{\rm noise}>0.015$ and in $\sigma_{\rm noise}=0.023$, when $\Delta\,10\%$ of \vsinitheo~($270 \, \rm km \ s^{-1}$ in this case) is reached, then the velocity yielded is not reliable, given by MC simulations. This critical $\sigma_{\rm c}$ (or $\sigma_{\rm noise}$) is defined as $\sigma_{\rm c, crit}$ (or $\sigma_{\rm noise, crit}$). For the case of FEROS resolution, the decrease velocities is at higher noise, being $\sigma_{\rm noise}=0.024$ when the median of the 10~000 MC is reached for 270 $\rm km \ s^{-1}$. In addition, error bars suggest that for FEROS is lower than PUCHEROS. For this reason an instrument with high resolution yields more reliable results.

\begin{figure}
\resizebox{8.5cm}{!}{\includegraphics{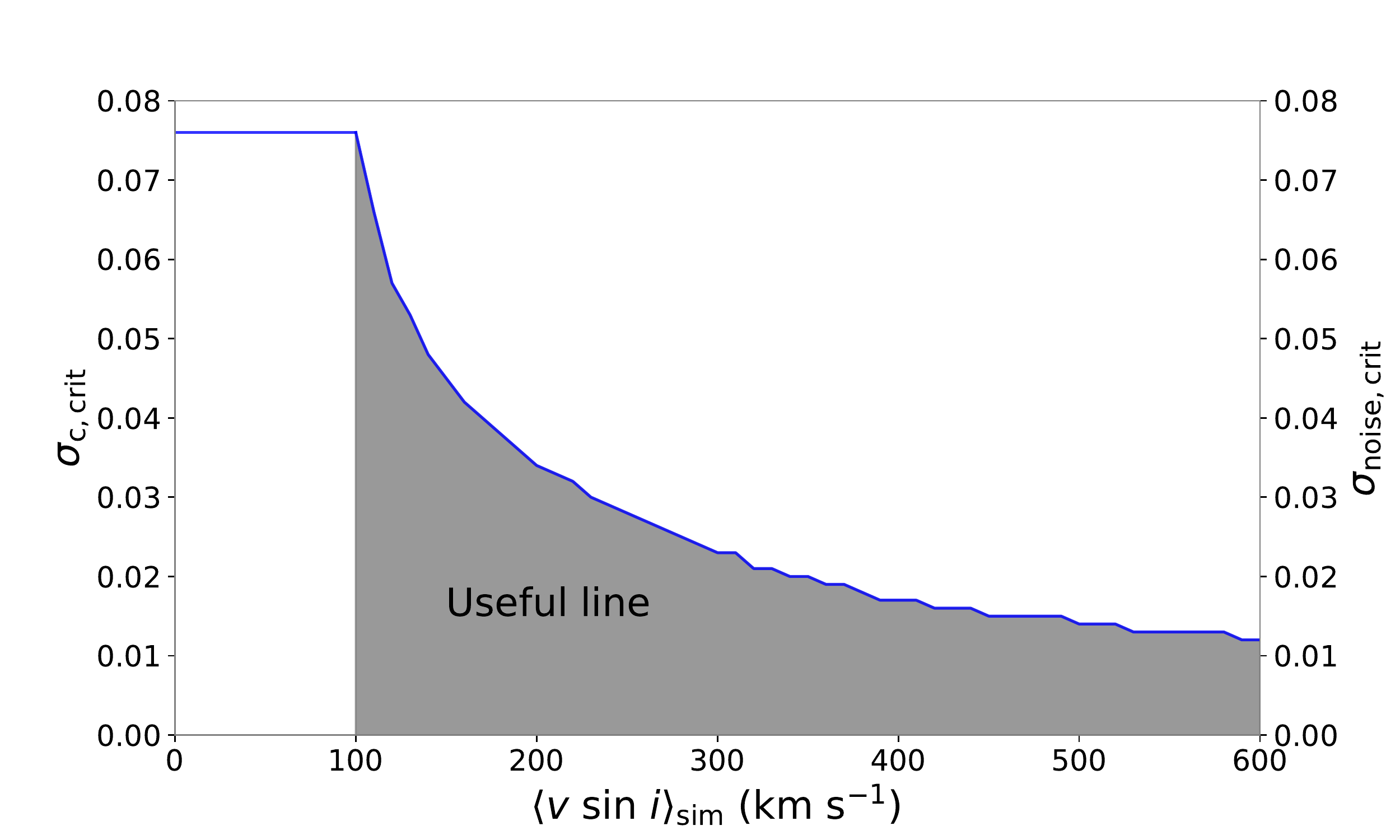}}
    \caption{Critical standard deviation of the continuum for observations ($\sigma_{\rm c, crit}$) or simulations ($\sigma_{\rm noise, crit}$) as a function of \vsini~via FT (solid blue curve). We notice that these values were obtained for a $\Delta \lambda =0.089$~\AxP~(PUCHEROS resolution). Values below 100 km s$^{-1}$ are unreliable.}
    \label{fig3.2}
\end{figure}

We adopt this threshold as our last criteria to select ``good'' lines, the sixth criterion, which yields an extra support to the procedure given our MC simulations and is presented in the last row of Table~\ref{tab:criteria}. Figure~\ref{fig3.2} shows the behaviour of $\sigma_{\rm c, crit}$ for the grid of simulated \vsini. For high rotators the critical measured noise $\sigma_{\rm c,crit}$ is lesser than slow rotators and this is because as rotational velocity increases, the magnitude of the signal decreases and is broadened, which makes the observed line harder to detect (same case of $\sigma_{\rm noise} = 0.01$ and \vsinitheo$= 400 \ \rm km \ s^{-1}$ in Fig.~\ref{fig2.1}).

It is important to mention that for slow rotators, the number of bins within a thin absorption line is insufficient to compute the FT. In the study of \cite{SimonDiaz2014} they concluded that any \vsini~measurement below $40 \ \rm km \ s^{-1}$ using FT has to be taken as the upper limit (if effects of microturbulence are not included). The data from that work were given by the FIbre-fed Echelle Spectrograph (FIES) medium and low- resolution with resolutions of $R\,\sim$~23~000 and 46~000, respectively, both having a spectral dispersion of 0.025~\AxP. For our case, we use the BeSOS database which has the same bin steps for simulations implemented in this work. In our work, the simulation shows that values below 100\kps~and $\sigma_{\rm c,crit}$ are considered as unreliable results. However, as majority of CBes rotate higher than this velocity this is not a problem and this should be taken account for stars of lower rotation (see Table \ref{tab3}).

\section{Application to BeSOS database}\label{sec:5.1}
\begin{figure*}
\resizebox{14.95cm}{!}{\includegraphics{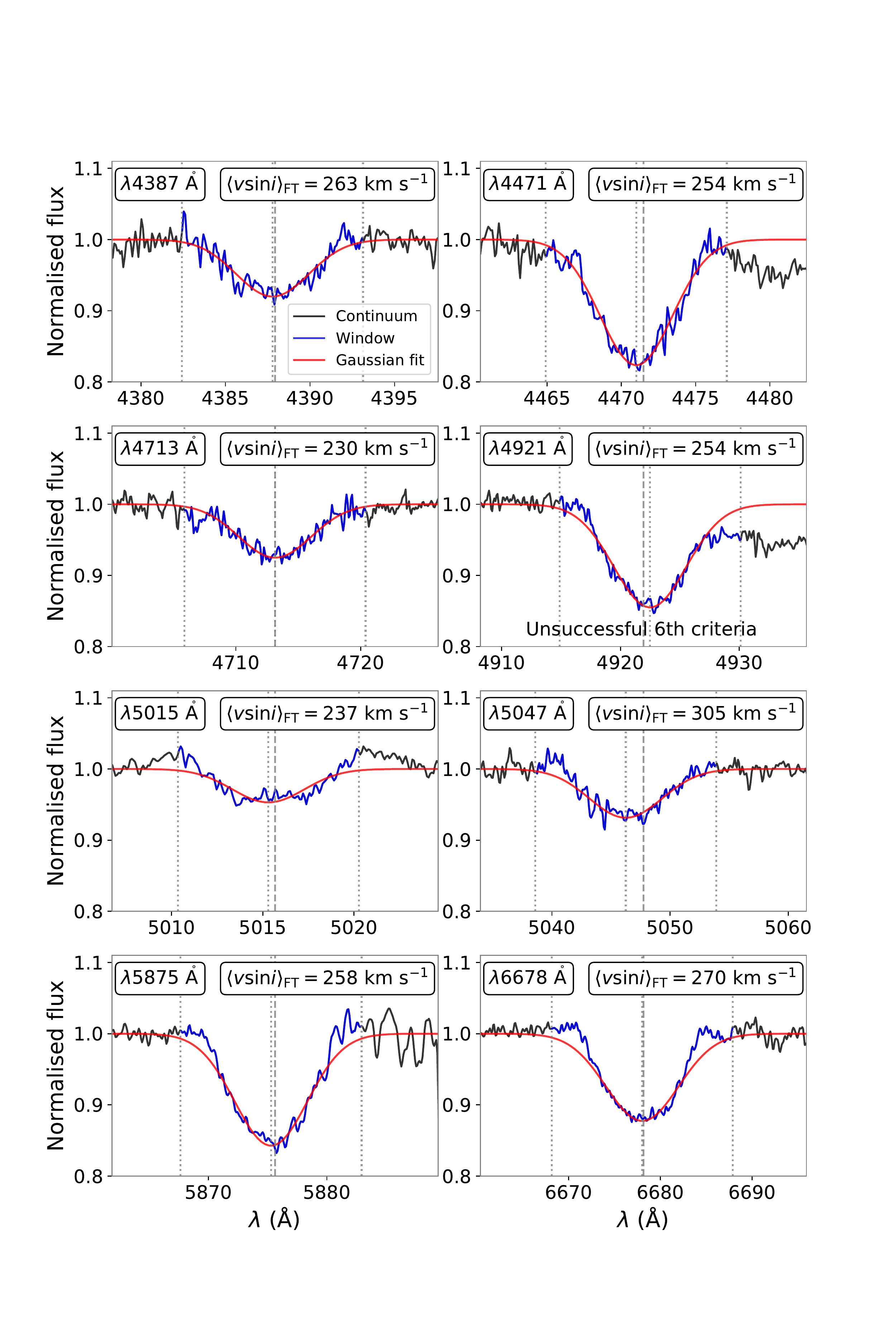}}
    \caption{Procedure applied to the eight \HeI~spectral lines in HD157246 for a window of 2.5~\sigmafit. Normalised flux as a function of wavelengths. Solid red curve represents the best Gaussian fitted, blue solid lines are the spectral signal selected and solid black lines the continuum. In the inset of each subplot the midwave and rotational speed obtained are showed. Vertical dotted grey lines represents the limit of the signal and midwave given by the Gaussian fit. Vertical dashed grey lines are the laboratory \HeI~midwave. $\lambda$4921~did not satisfy the sixth criterion.}
    \label{fig4.1}
\end{figure*}

The procedure was applied to BeSOS database for 64 stars at different epochs with a total of 314 spectra.
As we use eight \HeI~absorption lines, in order to visualise the spectra, Fig. ~\ref{fig4.1} shows the procedure implemented to HD157246 for a single epoch using a window of 2.5~\sigmafit.

\begin{figure}
\resizebox{8.5cm}{!}{\includegraphics{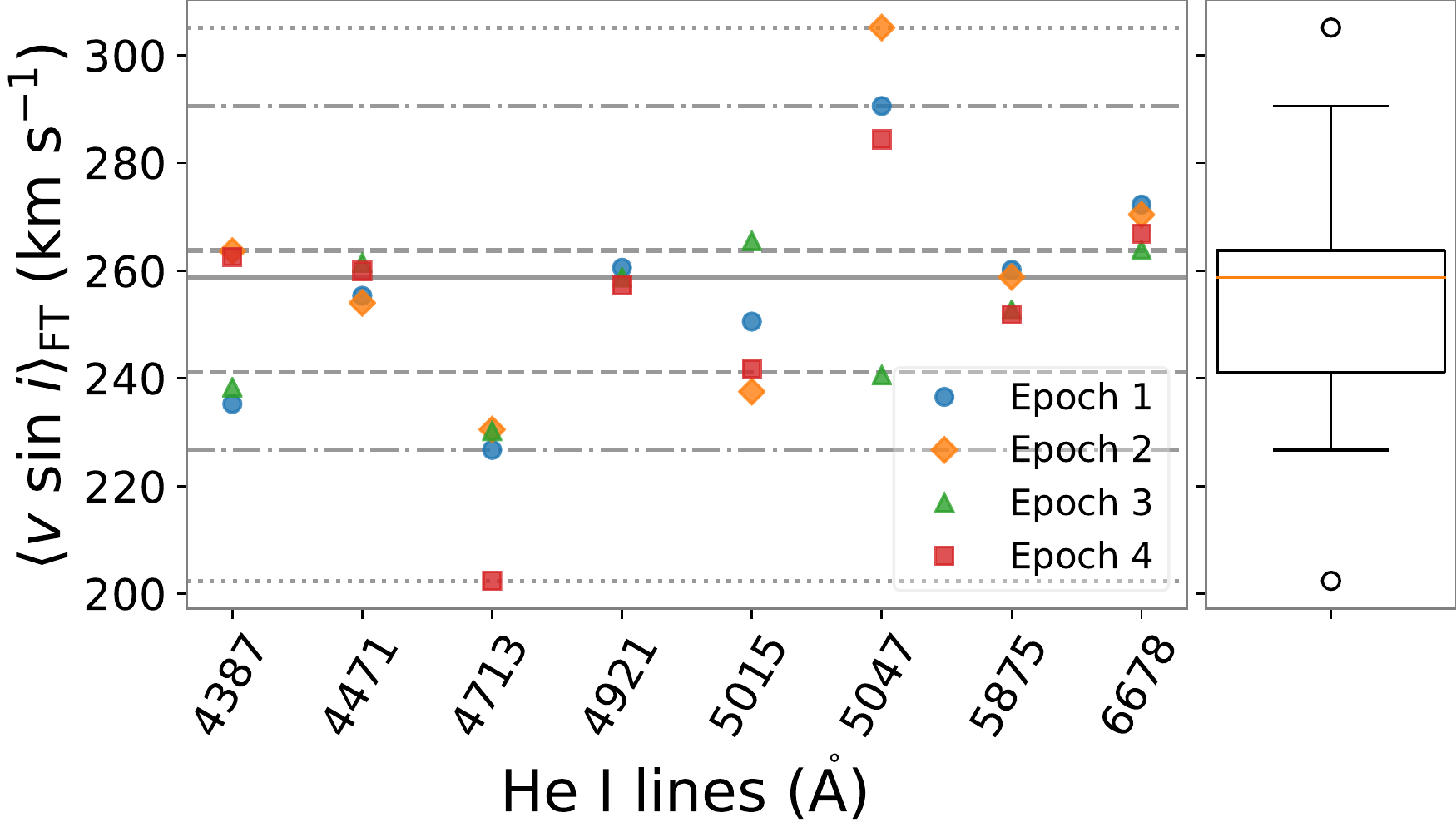}}
    \caption{Projected rotational velocities obtained for HD157246 at different epochs as a function of the respective \HeI~line. Left plot: Each symbol and colour is a specific spectrum for an unique epoch. These epochs (1 - 4) are given by 56~735.3323, 57~078.3089, 57~218.1718, 57~218.1793 in Modified Julian Dates, respectively. The horizontal solid, dashed, dashed-dotted and dotted grey lines gives the median, first and third quartile, minimum and maximum, and outliers, respectively. Right panel: Standard boxplot for all \vsini~values given by the different epochs.}
    \label{fig4.2}
\end{figure}

In this figure, seven of eight lines were used to measure \vsini~because in case of $\lambda$4921, the sixth criterion detect a high level of noise in the continuum. This last case is interpreted by different normalised continuum rates in the left and right sides from the midwave, values of $\sim 1$ and $\sim 0.95$, respectively. Non-radial pulsation and stellar winds for Be stars tend to be the main source of asymmetry, but it is not discarded the influence of other processes such as line blendings with near transitions, differential rotation or even minor physical phenomena such as turbulence. However, we tried to select only helium line profiles that were less affected by asymmetries for our study using the sixth criterion. \cite{SimonDiaz2014} presents an friendly-interactively code that solves this problem selecting the continuum part manually. In Fig.~\ref{fig4.2}, the \vsini~values are represented for HD157246 at four different epochs.

To determine the average per star boxplots for all \vsiniFT~were computed. This technique consist in obtain the first quartile, the median and the third quartile, and the interquartile range (IQR). Following this, to exclude outliers the 1.5 IQR above and below the third and first quartiles are calculated. The values above or below this range are considered as outliers. As can be noted the dispersion of values is still high; it changes from $200 \ {\rm km \ s^{-1}} < \,$\vsini $< 310 \ {\rm km \ s^{-1}}$ to $220 \ {\rm km \ s^{-1}} < \,$\vsini $\,< 290 \ {\rm km \ s^{-1}}$. Looking for a relation between the Modified Julian Date\footnote{$\text{MJD} = \text{JD} - 2400000.5$} (MJD) and the type of absorption line used in this star, it can be infer that there is no clear relation.

To survey the averaged \vsini~values for windows of 2.0~\sigmafit, 2.5~\sigmafit and 3.0~\sigmafit, Table~\ref{tab3} shows the results compared to \cite{Arcos2018} as ``Arcos+2018'' and \cite{Fremat2005} as ``Fremat2005'' were they calculated $v \sin i$ using synthetic spectra with classical plane parallel non-LTE model atmospheres using apparent fundamental parameters derived from \HeI~4471 line.

\begin{table*}
\centering
\begin{tabular}{llllcrr}
\toprule
Target & 2.0~\sigmafit & 2.5~\sigmafit & 3.0~\sigmafit &  N$^{o}$ of epochs & Arcos+2018 & Fremat2005 \\
\midrule
HD10144  &   228\scriptsize$_{-13}^{+31}$ &   225\scriptsize$_{-51}^{+29}$ &    225\scriptsize$_{-39}^{+59}$ &                        2 & ----- & ----- \\
HD33328  &   283\scriptsize$_{-50}^{+40}$ &   279\scriptsize$_{-52}^{+32}$ &    292\scriptsize$_{-30}^{+25}$ &                        5 & 287 & 333 \\
HD35165  &   213\scriptsize$_{-22}^{+63}$ &     217\scriptsize$_{-9}^{+4}$ &    211\scriptsize$_{-61}^{+36}$ &                        7 & 240 & 350 \\
HD35411  &  320\scriptsize$_{-76}^{+103}$ &   327\scriptsize$_{-44}^{+35}$ &    325\scriptsize$_{-86}^{+90}$ &                        5 & 53 & 174 \\
HD35439  &   244\scriptsize$_{-82}^{+60}$ &   230\scriptsize$_{-68}^{+77}$ &    253\scriptsize$_{-63}^{+61}$ &                        8 & 266 & 266 \\
HD37041  &   149\scriptsize$_{-43}^{+30}$ &   132\scriptsize$_{-16}^{+24}$ &    135\scriptsize$_{-14}^{+19}$ &                        5 & 140 & 133 \\
HD37795  &   190\scriptsize$_{-33}^{+27}$ &   181\scriptsize$_{-51}^{+23}$ &    188\scriptsize$_{-56}^{+94}$ &                        9 & ----- & 192 \\
HD41335  &   242\scriptsize$_{-64}^{+85}$ &   258\scriptsize$_{-86}^{+54}$ &   244\scriptsize$_{-71}^{+100}$ &                        8 & 330 & 376 \\
HD42167  &   199\scriptsize$_{-73}^{+74}$ &   203\scriptsize$_{-58}^{+47}$ &    190\scriptsize$_{-48}^{+52}$ &                        6 & ----- & 249 \\
HD45725  &   340\scriptsize$_{-19}^{+10}$ &   337\scriptsize$_{-20}^{+20}$ &    345\scriptsize$_{-15}^{+26}$ &                        2 & 280 & 345 \\
HD45910  &     196\scriptsize$_{-7}^{+6}$ &   131\scriptsize$_{-36}^{+59}$ &    137\scriptsize$_{-43}^{+56}$ &                        6 & 100 & 254 \\
HD48917  &   180\scriptsize$_{-30}^{+29}$ &   167\scriptsize$_{-41}^{+44}$ &    191\scriptsize$_{-53}^{+24}$ &                       10 & 200 & 212 \\
HD50013  &   231\scriptsize$_{-55}^{+34}$ &   209\scriptsize$_{-38}^{+49}$ &    228\scriptsize$_{-37}^{+30}$ &                        7 & 290 & 244 \\
HD52918  &   305\scriptsize$_{-24}^{+17}$ &    303\scriptsize$_{-14}^{+8}$ &     309\scriptsize$_{-4}^{+14}$ &                        2 & 242 & 265 \\
HD56014  &   183\scriptsize$_{-66}^{+25}$ &    199\scriptsize$_{-50}^{+6}$ &    129\scriptsize$_{-41}^{+45}$ &                        2 & 200 & 294 \\
HD57150  &   213\scriptsize$_{-18}^{+13}$ &    218\scriptsize$_{-23}^{+8}$ &     228\scriptsize$_{-19}^{+9}$ &                        2 & 180 & 190 \\
HD57219  &     135\scriptsize$_{-0}^{+0}$ &      88\scriptsize$_{-8}^{+7}$ &      84\scriptsize$_{-8}^{+12}$ &                        4 & 50 & 84 \\
HD58715  &   284\scriptsize$_{-86}^{+61}$ &  299\scriptsize$_{-94}^{+117}$ &   278\scriptsize$_{-66}^{+162}$ &                        3 & ----- & 231 \\
HD60606  &   249\scriptsize$_{-34}^{+29}$ &   256\scriptsize$_{-24}^{+25}$ &    258\scriptsize$_{-25}^{+23}$ &                        6 & 250 & 285 \\
HD63462  &   238\scriptsize$_{-47}^{+47}$ &   279\scriptsize$_{-69}^{+23}$ &    310\scriptsize$_{-78}^{+39}$ &                        3 & 300 & 514 \\
HD68980  &     152\scriptsize$_{-3}^{+0}$ &   127\scriptsize$_{-11}^{+18}$ &     141\scriptsize$_{-6}^{+10}$ &                        3 & 110 & 152  \\
HD71510  &   165\scriptsize$_{-28}^{+20}$ &   155\scriptsize$_{-30}^{+20}$ &    155\scriptsize$_{-23}^{+15}$ &                        6 & 150 & 154 \\
HD75311  &   249\scriptsize$_{-39}^{+34}$ &   247\scriptsize$_{-76}^{+38}$ &    256\scriptsize$_{-67}^{+21}$ &                        2 & 250 & 283 \\
HD78764  &   150\scriptsize$_{-36}^{+19}$ &   113\scriptsize$_{-23}^{+12}$ &    122\scriptsize$_{-22}^{+19}$ &                        3 & 140 & 120 \\
HD83953  &   286\scriptsize$_{-24}^{+69}$ &   263\scriptsize$_{-22}^{+24}$ &    257\scriptsize$_{-19}^{+25}$ &                        1 & 250 & 276 \\
HD89080  &   224\scriptsize$_{-45}^{+31}$ &   233\scriptsize$_{-29}^{+54}$ &   245\scriptsize$_{-63}^{+187}$ &                        7 & ----- & 254 \\
HD89890  &    130\scriptsize$_{-37}^{+4}$ &     92\scriptsize$_{-1}^{+46}$ &      97\scriptsize$_{-22}^{+2}$ &                        5 & 26 & 67 \\
HD91465  &   286\scriptsize$_{-19}^{+14}$ &   288\scriptsize$_{-20}^{+14}$ &    290\scriptsize$_{-14}^{+15}$ &                        6 & 280 & 285 \\
HD92938  &   131\scriptsize$_{-21}^{+37}$ &    121\scriptsize$_{-16}^{+9}$ &     122\scriptsize$_{-9}^{+16}$ &                        7 & 110 & 139 \\
HD93563  &  183\scriptsize$_{-54}^{+101}$ &   194\scriptsize$_{-55}^{+95}$ &    183\scriptsize$_{-48}^{+91}$ &                        6 & 280 & 275 \\
HD98058  &   223\scriptsize$_{-26}^{+52}$ &     201\scriptsize$_{-9}^{+7}$ &      219\scriptsize$_{-8}^{+2}$ &                        3 & ----- & 254 \\
HD102776 &   222\scriptsize$_{-40}^{+16}$ &   190\scriptsize$_{-25}^{+35}$ &    197\scriptsize$_{-25}^{+24}$ &                        5 & 200 & 270 \\
HD103192 &   275\scriptsize$_{-49}^{+11}$ &   275\scriptsize$_{-41}^{+34}$ &   274\scriptsize$_{-97}^{+251}$ &                        5 & ----- & 42 \\
HD105382 &    171\scriptsize$_{-5}^{+12}$ &  183\scriptsize$_{-108}^{+10}$ &    117\scriptsize$_{-17}^{+61}$ &                        5 & 67 & 72 \\
HD105435 &   257\scriptsize$_{-85}^{+34}$ &   245\scriptsize$_{-63}^{+35}$ &    245\scriptsize$_{-59}^{+46}$ &                        3 & 250 & 163 \\
HD107348 &   234\scriptsize$_{-52}^{+28}$ &   204\scriptsize$_{-31}^{+42}$ &    237\scriptsize$_{-43}^{+27}$ &                        5 & ----- & 237 \\
HD110335 &   242\scriptsize$_{-75}^{+15}$ &   214\scriptsize$_{-23}^{+34}$ &    216\scriptsize$_{-13}^{+29}$ &                        3 & ----- & ----- \\
HD110432 &    181\scriptsize$_{-3}^{+52}$ &    184\scriptsize$_{-7}^{+22}$ &      185\scriptsize$_{-2}^{+2}$ &                        4 & 400 & 419 \\
HD112078 &    307\scriptsize$_{-2}^{+26}$ &    303\scriptsize$_{-75}^{+1}$ &    338\scriptsize$_{-34}^{+72}$ &                        1 & 290 & 327 \\
HD120324 &   151\scriptsize$_{-25}^{+13}$ &   129\scriptsize$_{-32}^{+21}$ &     139\scriptsize$_{-11}^{+8}$ &                        3 & 110 & 162 \\
HD124195 &   135\scriptsize$_{-34}^{+39}$ &   136\scriptsize$_{-26}^{+30}$ &    185\scriptsize$_{-53}^{+11}$ &                        2 & 150 & ----- \\
HD124367 &   248\scriptsize$_{-16}^{+16}$ &    272\scriptsize$_{-1}^{+16}$ &     277\scriptsize$_{-18}^{+5}$ &                        2 & 260 & 318 \\
HD124771 &   188\scriptsize$_{-15}^{+10}$ &   164\scriptsize$_{-35}^{+21}$ &    175\scriptsize$_{-20}^{+20}$ &                        2 & 150 & 205 \\
HD127972 &   245\scriptsize$_{-84}^{+47}$ &   203\scriptsize$_{-46}^{+57}$ &    206\scriptsize$_{-47}^{+69}$ &                        4 & 240 & 326 \\
HD131492 &   173\scriptsize$_{-54}^{+43}$ &   168\scriptsize$_{-45}^{+50}$ &    152\scriptsize$_{-31}^{+56}$ &                        6 & 100 & 192 \\
HD135734 &   257\scriptsize$_{-88}^{+72}$ &  290\scriptsize$_{-109}^{+44}$ &   276\scriptsize$_{-100}^{+62}$ &                        7 & ----- & 282 \\
HD138769 &   159\scriptsize$_{-21}^{+20}$ &   102\scriptsize$_{-16}^{+74}$ &     91\scriptsize$_{-15}^{+25}$ &                        5 & 30 & 78 \\
HD142983 &   227\scriptsize$_{-52}^{+83}$ &   238\scriptsize$_{-34}^{+79}$ &    250\scriptsize$_{-17}^{+53}$ &                        6 & 370 & 407 \\
HD143275 &   204\scriptsize$_{-38}^{+47}$ &   184\scriptsize$_{-32}^{+23}$ &     203\scriptsize$_{-7}^{+11}$ &                        2 & 257 & 150 \\
HD148184 &   157\scriptsize$_{-45}^{+23}$ &   134\scriptsize$_{-22}^{+37}$ &    159\scriptsize$_{-35}^{+30}$ &                        5 & 150 & 151 \\
HD157042 &   310\scriptsize$_{-59}^{+45}$ &   299\scriptsize$_{-33}^{+66}$ &    303\scriptsize$_{-29}^{+52}$ &                        5 & 280 & 348 \\
HD157246 &    257\scriptsize$_{-19}^{+6}$ &    258\scriptsize$_{-21}^{+8}$ &     262\scriptsize$_{-17}^{+7}$ &                        4 & 230 & 230 \\
HD167128 &     154\scriptsize$_{-1}^{+2}$ &     186\scriptsize$_{-3}^{+2}$ &      203\scriptsize$_{-1}^{+2}$ &                        3 & 50 & 55 \\
HD205637 &   225\scriptsize$_{-40}^{+46}$ &   247\scriptsize$_{-65}^{+26}$ &    259\scriptsize$_{-43}^{+11}$ &                        2 & 230 & 238 \\
HD209014 &   208\scriptsize$_{-25}^{+52}$ &   195\scriptsize$_{-43}^{+80}$ &    195\scriptsize$_{-33}^{+95}$ &                        8 & ----- & 357 \\
HD209409 &  254\scriptsize$_{-105}^{+46}$ &   244\scriptsize$_{-93}^{+50}$ &    240\scriptsize$_{-32}^{+41}$ &                        7 & 350 & 282 \\
HD212076 &   148\scriptsize$_{-21}^{+25}$ &   142\scriptsize$_{-54}^{+55}$ &    120\scriptsize$_{-30}^{+76}$ &                        6 & 100 & 103 \\
HD212571 &   227\scriptsize$_{-78}^{+48}$ &   225\scriptsize$_{-73}^{+43}$ &    239\scriptsize$_{-79}^{+44}$ &                        7 & 215 & 233 \\
HD214748 &   182\scriptsize$_{-22}^{+41}$ &   185\scriptsize$_{-41}^{+28}$ &    184\scriptsize$_{-27}^{+45}$ &                        9 & ----- & 205 \\
HD217891 &   222\scriptsize$_{-53}^{+30}$ &   277\scriptsize$_{-71}^{+20}$ &  271\scriptsize$_{-112}^{+111}$ &                       10 & 90 & 100 \\
HD219688 &   225\scriptsize$_{-22}^{+24}$ &   228\scriptsize$_{-47}^{+15}$ &     230\scriptsize$_{-54}^{+5}$ &                        4 & ----- & 316 \\
HD221507 &     237\scriptsize$_{-6}^{+3}$ &   225\scriptsize$_{-23}^{+18}$ &     221\scriptsize$_{-21}^{+6}$ &                       10 & ----- & ----- \\
\bottomrule
\end{tabular}
\caption{Projected rotational velocities obtained by our work for windows of $2\sigma_{\rm fit}$, $2.5\sigma_{\rm fit}$ and $3\sigma_{\rm fit}$ (in $\rm km \ s^{-1}$; 2nd, 3rd and 4th columns, respectively) with confidence interval of $16\%-84\%$. 5th column represents the number of epochs. 6th and 7th columns are \vsini~(in $\rm km \ s^{-1}$)  values obtained from literature.}
\label{tab3}
\end{table*}

Besides, in Fig. \ref{figApp} the relation between the \vsini~values obtained by our method (using $k =$~2.0, 2.5 and 3.0) are compared with literature results (Fremat2005 and Arcos+2018).

\section{Discussion}

\subsection{Lower limit of \vsini}
PUCHEROS versus FEROS resolutions evidence from Fig.~\ref{fig3.1} that the capacity of the instrument is a very important feature because the higher the SNR, the smaller is the \vsini~lower limit value that this method delivers. Based on our procedure to obtain the projected rotational velocities for BeSOS database, three stars (HD13876, HD89890 and HD5721) presented a \vsini~lower than $100 \ \rm km \ s ^{-1}$ and therefore their projected rotational speeds are considered equal to $100 \ \rm km \ s^{-1}$. \cite{Sundqvist2013} and \cite{SimonDiaz2014} claim that, using FT and goodness-of-fit (GOF) methods, it is not possible to derive reliable projected rotation speeds when \vsini~$\lesssim 40-50$~\kps .

\subsection{\HeI~\LHeIOne~line}
\begin{figure*}
\resizebox{14.95cm}{!}{\includegraphics{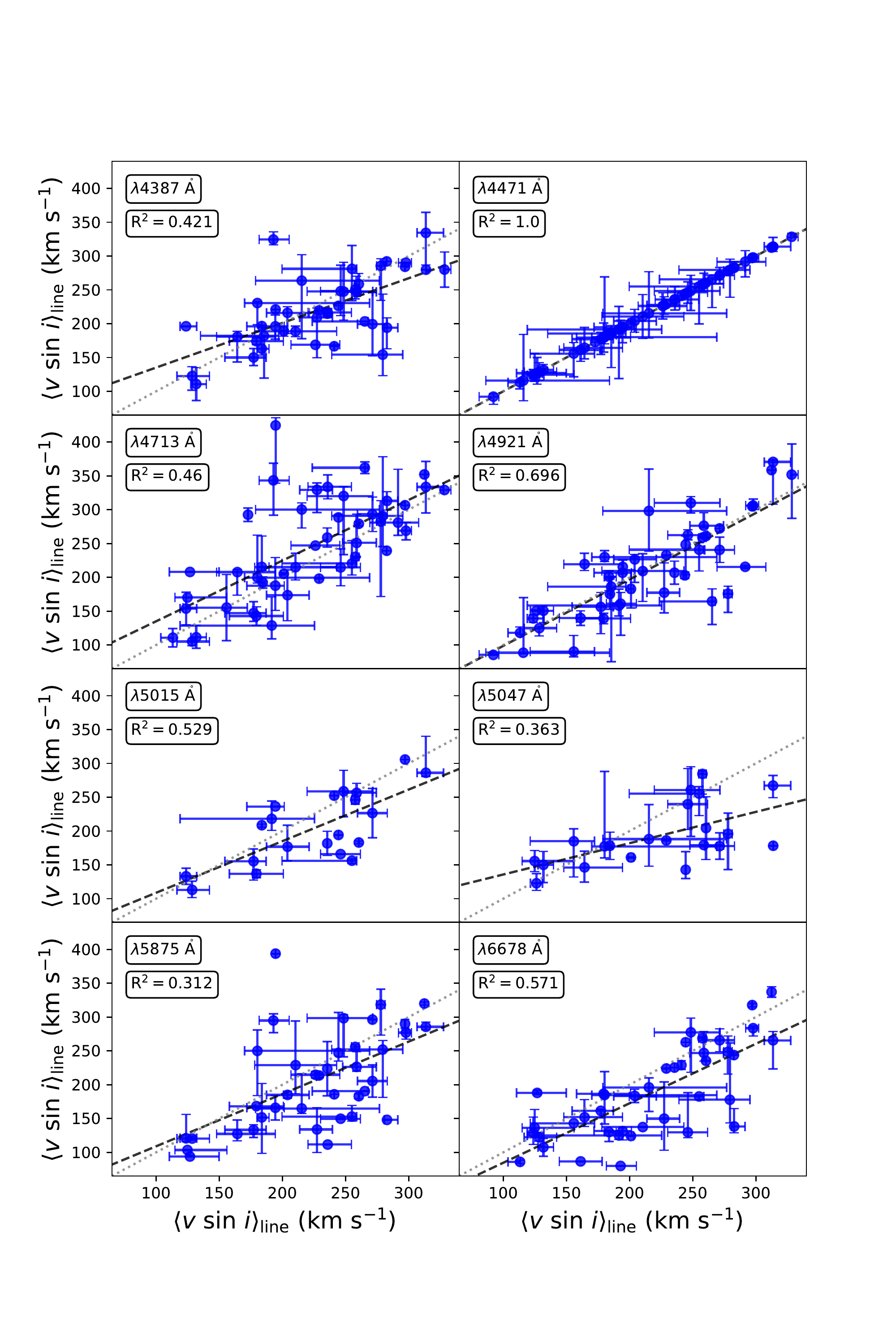}}
    \caption{Final \vsini~averaged per line (\vsiniline) as a function of \vsiniline~for \HeI~\LHeIOne~using a window of 2.5~\sigmafit (black dots). Error bars are given by a $16\%-84\%$ confidence interval. In the inset of each plot the \HeI~lines and the correlation coefficient squared are represented. The dashed black line are the least square polynomial fit. The 1:1 relation is also showed (dashed grey line).}
    \label{fig5.1}
\end{figure*}

Having the velocities averaged per line for each star, Fig.~\ref{fig5.1} shows the behaviour for \LHeIOne~\HeI~line versus all the other ones used ($\lambda\lambda$4387, 4713, 4921, 5015, 5047, 5875 and 6678) for a window of 2.5~\sigmafit and with confidence intervals of $16\%-84\%$. This absorption line is chosen for comparison reasons since it is essential to analyse how the other lines are approached with respect to \LHeIOne~\HeI~line.

Two fits are included in this figure, a linear least square polynomial fit and a 1:1 relation line in order to analyse relations for \vsini~values at different \lamtheo. Correlation coefficients are included on the insets for each subplot. The \LHeIOne~with the $\lambda$4921 lines are correlated very well with 0.696 and indeed, the 1:1 relation is very similar to the linear fit. Giving that in general, these two lines can be conceived as equal lines. The cases that are relatively well correlated are \LHeIOne~versus $\lambda$5015 and $\lambda$6678 with the difference that in both cases the velocities are slightly underestimated. For the following samples are $\lambda$4387 and $\lambda$4713, in which $\rm R^{2}$ are 0.421 and 0.46, respectively. On the other hand, the linear fits, compared to \LHeIOne, are underestimated (for $\lambda$4387) and very similar ($\lambda$4712).

\begin{figure*}
\resizebox{14.95cm}{!}{\includegraphics{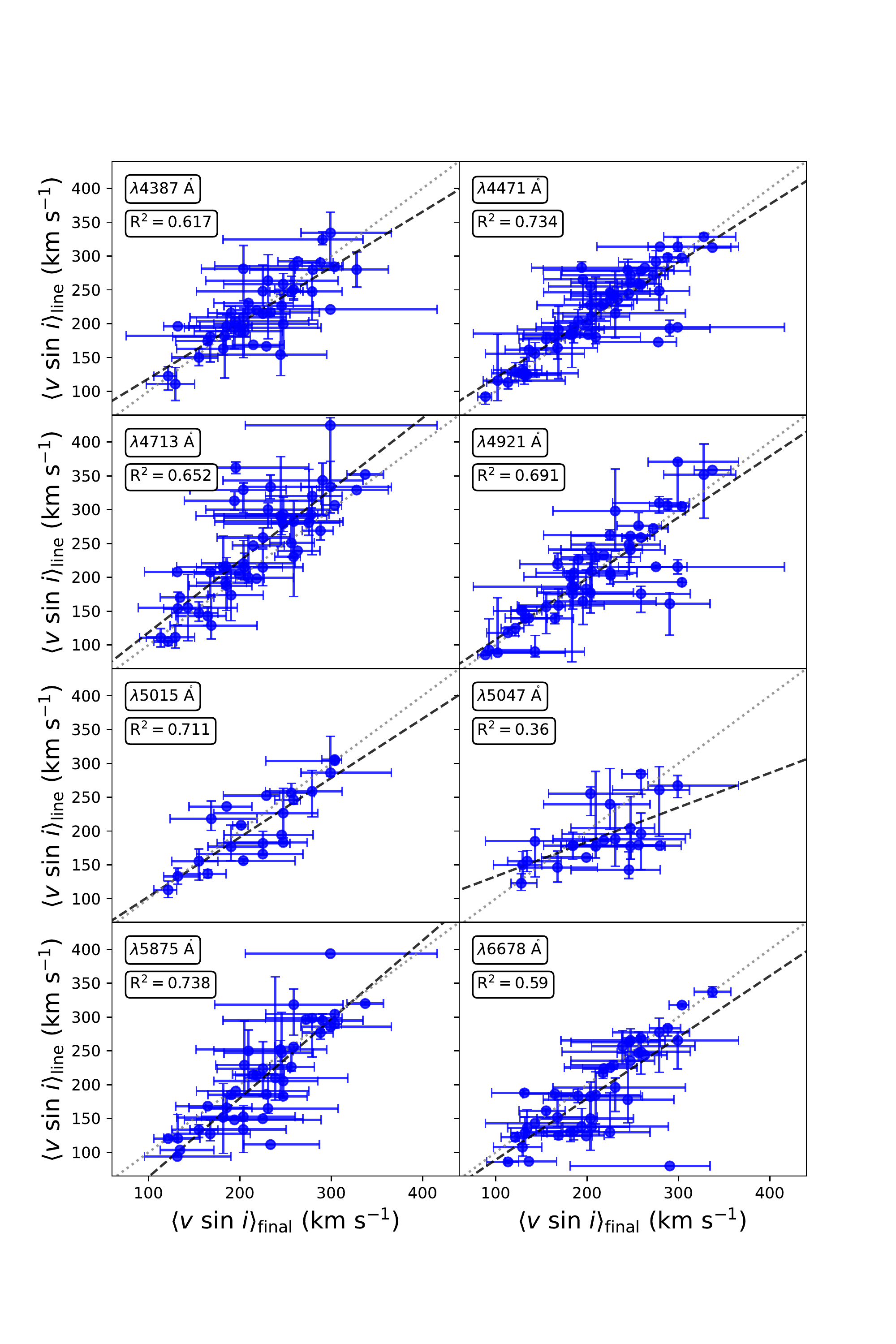}}
    \caption{\vsiniline~as a function of the final \vsini~averaged for the star using all \HeI~lines (\vsinifinal) with a window of $2.5\sigma_{\rm fit}$ (black dots). Error bars are given by a $16\%-84\%$ confidence interval. In the inset of each plot the \HeI~lines and the correlation coefficient squared are represented. The dashed black line are the least square polynomial fit. The 1:1 relation is also showed (dashed grey line).}
    \label{fig5.2}
\end{figure*}

As a general perspective, there is a trend for the other seven \HeI~lines ($\lambda \lambda$4387, 4713, 4921, 5015, 5047, 5875 and 6687) to underestimate \vsini~values with respect to the values from \LHeIOne. Alternatively, the average per star (same as Table~\ref{tab3}) is plotted as a function of the eight \vsiniline~in Fig.~\ref{fig5.2} for a window of 2.5~\sigmafit and with confidence intervals of $16\%-84\%$. 

In this figure, the best correlation factor ($R^{2}=0.738$) is for $\lambda$5875 with respect to $\lambda$4471. For lower velocities, in average, the values are underestimated and for higher velocities are overestimated, which yields in a fit upper the 1:1 relation. The cases of $\lambda\lambda$ 4471 and 5015 gives a squared correlation factor of 0.734 and 0.711, respectively, with the difference that now the fits are underestimated with respect to the 1:1 relation. The same happens with $\lambda\lambda$ 4387, 4921 and 6687 but with lower correlation factors. The other sample that overestimate with respect to \vsinifinal~is the $\lambda$4713 line with a good correlation. The scenario given by $\lambda$5047 with a squared degree of correlation factor of 0.36 and a fit that underestimate significantly the velocities. As a general perspective, the average of velocities obtained using our method, in comparison for \vsiniline~yields a varied range of correlations what can be said that results differs for different \HeI~lines to measure \vsini.

Despite the use of \HeI~lines, some of this variety of results could be due to the contamination of the disc to observe the photosphere on the stars. More studies in CBes, or massive high rotator stars, should be done at different epochs regarding the structure of a disc at different epochs to study if effectively this envelope contaminate the absorption line. Another theory is that not all the \HeI~must be distributed only on the equator, the poles, uniformly, and/or even, irregularly. Each transition could have a preferred location in the photosphere and deeper physics calculations should be done to improve this approach. Regarding to $\lambda$4437, this line was not taken into account for the 314 spectra analysed and it is an useless absorption line to measure \vsini~on CBes. This could be because is an optically thin line.

\subsection{The epsilon parameter}
For eq.~\ref{eq1} it is possible to display $\varepsilon$ as a function of the color index B - V ($I(\cos \theta)$), i. e., proportional to the temperature \citep{Gray2005}. The astronomical community since the work of \cite{Carroll1928} assume a LD coefficient $\varepsilon=0.6$ by default. This supposition is not quite true because the range of all spectral types stars, a difference of temperature is supported for separate stars. The expression of this coefficient is interpreted as an uniformly illuminated stellar disc for $\varepsilon=0$ to a disc fully darkened for $\varepsilon=1$. Furthermore, every single star should confine different features and is not useful to establish this value habitually. As a future work, we pretend to develop a grid of LD coefficients using non-linear models of LD laws \citep{Wade1985, Levenhagen2014} to fit the spectrum to well-known stars.

\subsection{Fast rotational effect}

The classical gravity darkening theory \citep{vonZeipel1924_a, vonZeipel1924_b} describes
the local dependency of temperatures and radiative fluxes with the fractional radius
of the star, which is dependent on the colatitude for a fast spinning star and gives rise to
an oblate geometry \citep{Harrington1968}. Although sounding reasonable, a detailed examination of the
classical von Zeipel's equation results in a problem known as the von Zeipel's paradox \citep{Zorec2017}, 
where stellar layers outside the stellar core present nonzero divergence of the radiative  
flux that gives rise to meridional circulations known as Eddington-Sweet currents \citep{Tassoul2007}. 
Further improvements to the theory were proposed \citep{Espinosa-Lara2011,Zorec2017} leading to a
different structure of fractional radii and a different dependence of temperatures and
radiative fluxes with the stellar colatitude.

\begin{figure}
\resizebox{8.5cm}{!}{\includegraphics{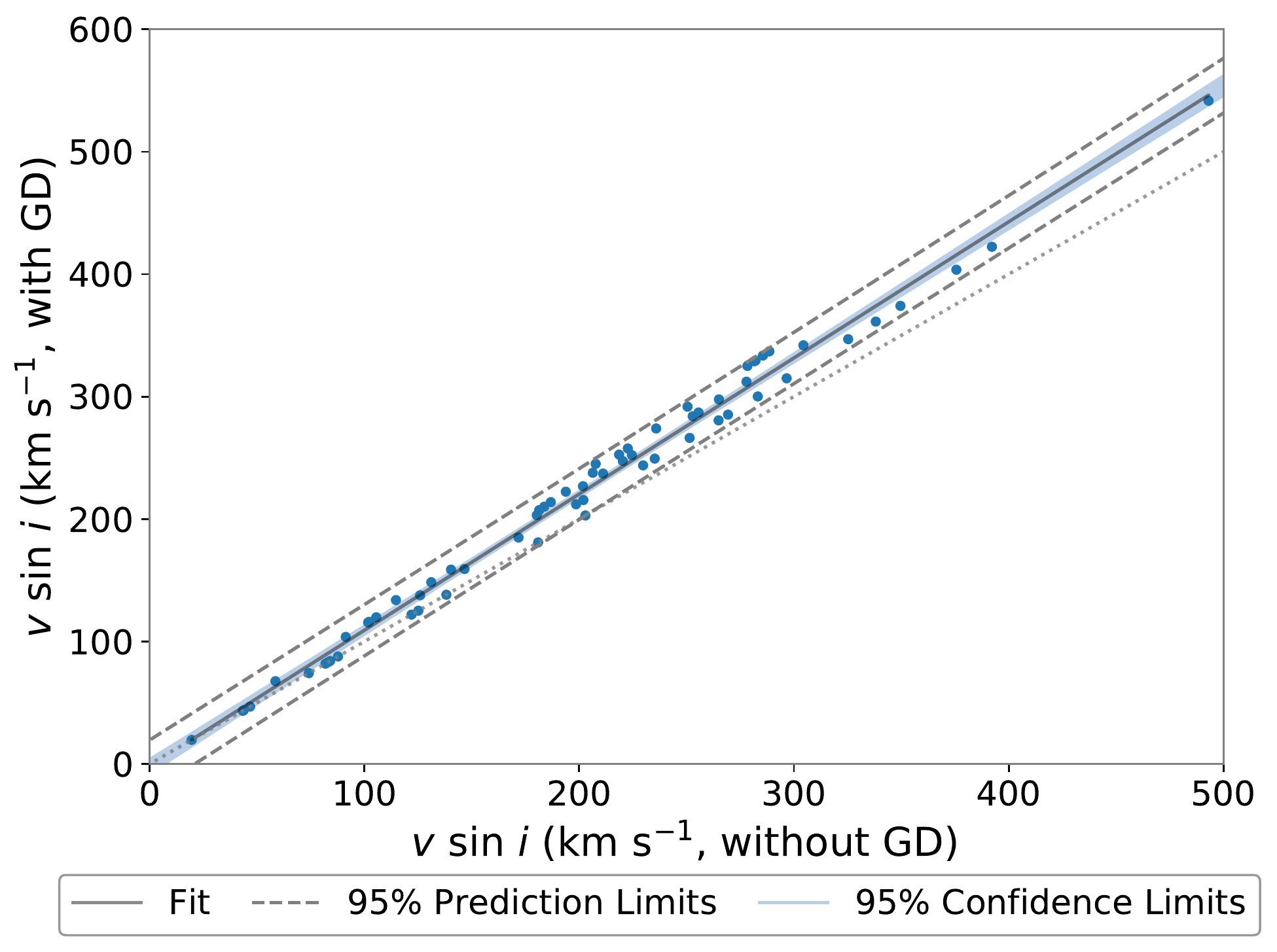}}
    \caption{\vsini~using GD as a function of \vsini~without GD models (blue dots). The dotted grey line is a 1:1 relation.}
    \label{fig5.3}
\end{figure}

Our FT analysis in this work considers spherical shapes and rigid rotation, neglecting subtle GD 
and geometrical deformation. It is worth emphasising that previous works revealed that projected 
rotational velocities measured from the fittings of classical model atmospheres  
or even FT techniques are indeed apparent quantities if these effects are not considered, 
leading possibly to \vsini~ underestimations \citep{Stoeckley1968,Townsend2004,Zorec2016,Zorec2017}. 
In order to figure out, to some extent, the difference between classical FT estimations 
and models accounting for GD, we performed fittings of GD models to the observed stellar 
spectra using a code to account for the classical von Zeipel's effect
\citep[ZPEKTR;][]{Levenhagen2014,Levenhagen2021}. The apparent \vsini~ values were
estimated with FT with a spectral window size of 3.0~\sigmafit~(Fig.~\ref{fig5.3}). The 
expected ``true'' \vsini~ values were derived through the use of correction factors $\delta$\teff,
$\delta$\logg~ and $\delta$\vsini~ by \cite{Fremat2005} to the averaged GD parameters obtained 
during the fitting procedures with ZPEKTR.

As can be noticed in the figure, modelling GD effects yield higher values and this is more considerable for velocities higher than 100 $\rm km \ s^{-1}$. Along with this result, a linear fit is computed to quantify how is the behaviour for this overestimation. The linear fit for \vsini~with GD versus without GD is given by $y = 1.111\,x - 2.013$ yielding a clearly underestimation of $\sim$10~\% for projected rotational speeds when GD effects are not taken into account. Confidence and prediction limits are also shown.

\section{Conclusions}

Stellar rotation is a fundamental parameter to constrain models of stellar formation and evolution. There is a wide field of research regarding stellar rotation: evolution and internal mechanism, GD, differential rotation, magnetic fields, angular momentum, convection/radiation energy transport, stellar oscillations, chemical mixing, mass-loss, rotation braking, among others \citep{Meynet2000}. In this work we presented a new procedure for the automatic \vsini~estimation of fast rotating Be-type stars. This procedure, based on the FT method proposed by \cite{Carroll1933_a}, showed to be suitable for stars with \vsini $> 50$~\kps~using a high-resolution instrument of $R\,\sim\,65 000$ \citep{Sundqvist2013}.
 
As Doppler effect broad intrinsic line profiles in atomic transitions for rotating stars, it is possible to measure \vsini~data via FT technique in high rotator stars (not only CBe) since broadening of lines are mainly dominated by rotation being the other effects neglectable, as microturbulence, macroturbulence, instrumental errors, among others. We performed a procedure that gives the \vsini~value for any selected absorption line profiles. In our case, the photosphere of CBes, are the lines \HeI~\AllLamdas. A Gaussian profile is fitted to the absorption line to select automatically the signal. The dispersion parameter of this Gaussian fit, \sigmafit, is used to determine the different windows sizes by $k$~\sigmafit.

We perform MC simulations for a grid of velocities (100 to 600 in steps of 10\kps) and synthetic Normal noise added (0.001 to 0.1 in steps of 0.001 from pseudo-random dispersion parameter). The MC simulations forced the foundation for a new sixth criteria in which it is necessary to measure the standard deviation of the continuum and if it is higher than a threshold from Fig.~\ref{fig3.1} therefore is not considered as a ``good'' line and is dropped.

The procedure was applied to BeSOS database for 64 stars at different epochs with a total of 314 spectra. The results are in global agreement in comparison to \cite{Arcos2018} and \cite{Fremat2005}, with the main difference that our method includes an average for eight \HeI~lines at different epochs. The importance of develop an automatic algorithm yields in the research field of Big Data Era and the fact that this method can include any midwave, being our case these eight \HeI~lines, a more detailed statistic research could be done for rotational velocities in high rotators and/or, even so, slow rotators at high resolution spectra for large databases. This study would conceive a new branch of the astrophysics in stellar rotations.

As \HeI~\LHeIOne~is the most common absorption line to measure \vsini~in CBes \citep{Zorec2016}, we have extended the survey by analysing other \HeI~lines: \AllLamdas. In summary, measuring the FT signal for different lines gives non-similar values.

For a complete study, non-linear solutions should be taken into account using a grid of limb-darkening coefficients to simulate the rotation profiles of stellar photospheres. Also, effects of gravity darkening need to be included \citep{Fremat2005,Zorec2017} since for fast rotators, the 
emerging flux from equatorial regions are reduced as the local gravity in these regions are small, which leads to the underestimation of
\vsini. We plan to employ the ZPEKTR code \citep{Levenhagen2014} to include the classical GD effects and also the Espinosa-Lara's scheme \citep{Espinosa-Lara2011} for high rotating stars in further studies.

\section*{Acknowledgements}

M.S. acknowledges grant support from project Beca de Magíster Nacional 2020, ANID (ex Conicyt), folio 22200219. M.C. and C.A. acknowledges partial support from Centro de Astrofísica de Valparaíso. MC also acknowledge support from Centro Interdiciplinario de Estudios Atmosfericos y Astroestadística, Universidad de Valparaíso. This work used BeSOS Catalogue, operated by the Instituto de Física y Astronomía, Universidad de Valparaíso, Chile: http://besos.ifa.uv.cl and funded by Fondecyt iniciación N° 11130702. C.A. thanks to FONDECYT N° 11190945. C.A., M.C. and I.A. thank the support from FONDECYT project N.1130485. I.A. thanks the support from FONDECYT project 11190147. M.S., M.C. I.A. and C.A. thank to project ANID-FAPESP Nº133541. This work has been possible thanks to the use of AWS-U.Chile-NLHPC credits. Powered@NLHPC: This research was partially supported by the supercomputing infrastructure of the NLHPC (ECM-02). This project has received funding from the European Union’s Framework Programme for Research and Innovation Horizon 2020 (2014-2020) under the Marie Skłodowska-Curie Grant Agreement No. 823734.

\textit{Softwares:} \textsc{Matplotlib} \citep{Hunter2007}, \textsc{AstroML} \citep{Vanderplas2012}, \textsc{ZPEKTR} \citep{Levenhagen2014}, \textsc{AstroPy} \citep{Astropy2018}, \textsc{SciPy} \citep{Virtanen2020}

\section*{Data Availability}
The data underlying this article will be shared on  request to the corresponding author

\bibliographystyle{mnras}
\bibliography{bibliography}

\appendix

\section{Velocities obtained in our work compared with literature}\label{Append}

\begin{figure*}
\resizebox{14.95cm}{!}{\includegraphics{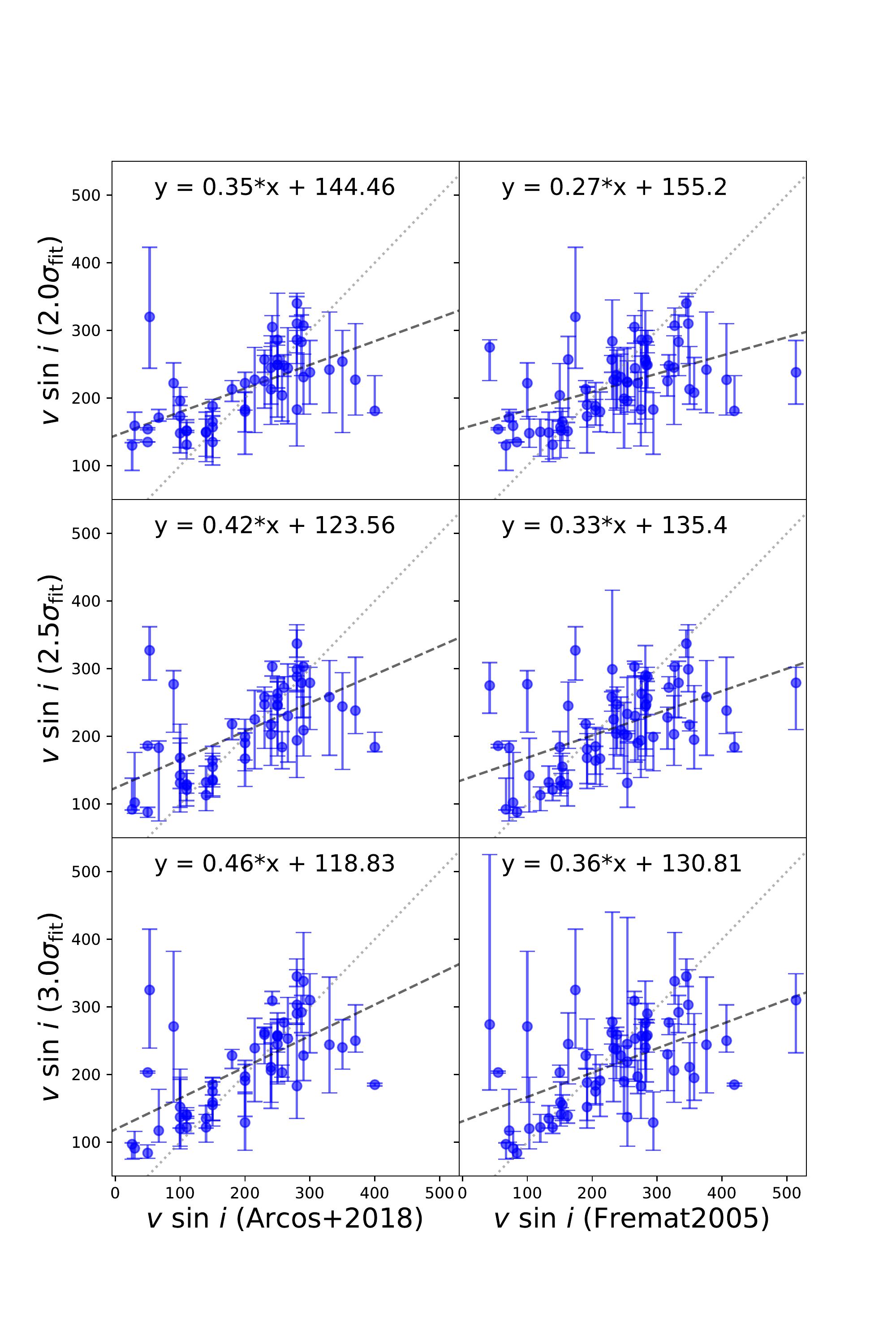}}
    \caption{\vsini~values obtained by our method (errorbars are percentile 16th and 84th) as a function of projected rotational velocities literature results in blue dots. Top, middle and bottom panels are for $k=$ 2.0, 2.5 and 3.0 window sizes, respectively. Left and right panels corresponds to Arcos+2018 and Fremat2005, respectively. Dashed black and dotted grey lines are the linear fit to dots and 1:1 relation.}
    \label{figApp}
\end{figure*}

In order to compare the values of Table \ref{tab3}, in Fig. \ref{figApp} there is shown the relation of these results.

In general terms, it can be inferred that for slow rotators ($\lesssim$ 200 km s$^{-1}$) our values overestimate the ones from Arcos+2018 and Fremat2005 and for high rotators ($\gtrsim$ 200 km s$^{-1}$) our values are underestimated, where if GD effects are taken into account then the results are improved.

%%%%%%%%%%%%%%%%%%%%%%%%%%%%%%%%%%%%%%%%%%%%%%%%%%
% Don't change these lines
\bsp	% typesetting comment
\label{lastpage}
\end{document}